\documentclass[12pt,preprint]{aastex}

\usepackage{amsmath}
\usepackage{natbib}
\usepackage{graphicx}
\usepackage{color}
\usepackage[breaklinks,colorlinks,citecolor=blue]{hyperref}
\usepackage{epsfig}
\usepackage{epstopdf}

\begin{document}

\title{Black Hole Space-time In Dark Matter Halo}

\author{
  Zhaoyi Xu,
  Xian Hou,
  Xiaobo Gong,
  Jiancheng Wang 
 }

\altaffiltext{1}{Yunnan Observatories, Chinese Academy of Sciences, 396 Yangfangwang, Guandu District, Kunming, 650216, P. R. China; {\tt zyxu88@ynao.ac.cn,xianhou.astro@gmail.com,xbgong@ynao.ac.cn,jcwang@ynao.ac.cn}}
\altaffiltext{2}{University of Chinese Academy of Sciences, Beijing, 100049, P. R. China}
\altaffiltext{3}{Key Laboratory for the Structure and Evolution of Celestial Objects, Chinese Academy of Sciences, 396 Yangfangwang, Guandu District, Kunming, 650216, P. R. China}
\altaffiltext{4}{Center for Astronomical Mega-Science, Chinese Academy of Sciences, 20A Datun Road, Chaoyang District, Beijing, 100012, P. R. China}

%\shorttitle{Black Hole Space-time In Dark Matter Halo}
\shortauthors{Z Y. Xu et al.}

\begin{abstract} 
For the first time, we obtain the analytical form of black hole space-time metric in dark matter halo for the stationary situation. Using the relation between the rotation velocity (in the equatorial plane) and the spherical symmetric space-time metric coefficient, we obtain the space-time metric for pure dark matter. By considering the dark matter halo in spherical symmetric space-time as part of the energy-momentum tensors in the Einstein field equation, we then obtain the spherical symmetric black hole solutions in dark matter halo. Utilizing Newman-Jains method, we further generalize spherical symmetric black holes to rotational black holes. As examples, we obtain the space-time metric of black holes surrounded by Cold Dark Matter and Scalar Field Dark Matter halos, respectively. Our main results regarding the interaction between black hole and dark matter halo are as follows: (i) For both dark matter models, the density profile always produces ``cusp" phenomenon in small scale in the relativity situation; (ii) Dark matter halo makes the black hole horizon to increase but the ergosphere to decrease, while the magnitude is small; (iii) Dark matter does not change the singularity of black holes. These results are useful to study the interaction of black hole and dark matter halo in stationary situation. Particularly, the ``cusp" produced in the $0\sim 1$ kpc scale would be observable in the Milky Way. Perspectives on future work regarding the applications of our results in astrophysics are also briefly discussed.
\end{abstract}


\section{INTRODUCTION} 
The origin of supermassive black hole in the center of galaxies is an important problem for high energy astrophysics. The dark matter halo may be helpful for us to solve this problem in the early universe \citep{2002PhRvL..88j1301B,2002ApJ...568..475B}. For general situation, the growth of black hole in dark matter halo with time has been analyzed through numerical simulations, but the analytical form of  black holes surrounded by dark matter halo has not yet been obtained. The reason is that when the interaction between dark matter and black hole is considered, the behaviour of the dark matter particle is unknown and fuzzy. Unfortunately, for stationary situation the effect of dark matter halo on the black hole is not obtained either analytically. For the Navarro-Frenk-White (NFW) dark matter model, the space-time geometry without black hole has been obtained and generally discussed \citep[See e.g., ][]{2003astro.ph..3594M,2004A&A...413..799F,2005GReGr..37..769M}. In their work, they assume that the pure dark matter space-time geometry is ``almost flat" because the dark matter density is small and no relativistic motion appears. Using these results, one could discuss several dynamical processes through geometric method when considering only pure dark matter halo, such as the tidal disruption effect (TDE), the motion of stars in dark matter halo, etc. If considering both dark matter halo and black hole, because of the lack of the space-time metric of black hole surrounded by dark matter halo, the above physical processes can not be studied in analytic way.

For a supermassive black hole in the center of a galaxy, its strong gravity could enhance the dark matter density significantly, producing a phenomenon named ``Spike'' \citep{1999PhRvL..83.1719G,2013PhRvD..88f3522S,2014PhRvL.113o1302F}. But for the NFW density profile, a ``cusp''  problem occurs \citep{2010AdAst2010E...5D}, and is contrary to observations which show rather a flat density profile. For other dark matter models, such as scalar field dark matter, modified newtonian dynamics dark matter and warm dark matter, ``cusp" is not produced in small scale. Whether the ``Spike" and ``Cusp" appear in the galactic center is unknown.

These problems inspire us to study black hole (spherical symmetric and rotational) in dark matter halo for stationary situation. Based on our results, we can investigate many dynamical processes near the black hole and the energy density of dark matter in relativistic limits.
%, we find that the cusp be always produced.

The paper is organized as follows. In Section 2, we introduce dark matter density profile and derive spherical symmetric dark matter space-time metric. In Section 3, we develop general method to study spherical symmetric black hole space-time surrounded by dark matter halo. In Section 4, we generalize spherical symmetric black hole to rotational black hole surrounded by dark matter halo. In Section 5, we discuss the behaviour of dark matter profile near the black hole and properties of black hole surrounded by dark matter halo. The summary is given in Section 6.

\section{DARK MATTER DENSITY PROFILE AND SPHERICAL SYMMETRIC DARK MATTER SPACE-TIME METRIC}
In this section, we derive the space-time geometry for pure dark matter. We follow the method introduced by \cite{2003astro.ph..3594M}, in which the spherical symmetric space-time metric with pure dark matter is given by
\begin{equation}
ds^{2}=-f(r)dt^{2}+\dfrac{dr^{2}}{g(r)}+r^{2}(d\theta^{2}+sin^{2}\theta d\phi^{2}).
\label{PDM1}
\end{equation}
For a test particle in spherical symmetric space-time, its rotation velocity in the equatorial plane is determined by the metric coefficient function $f(r)$ as
\begin{equation}
V^{2}=\dfrac{r}{\sqrt{f(r)}}\dfrac{d\sqrt{f(r)}}{dr}=r\dfrac{dln\sqrt{f(r)}}{dr}.
\label{PDM2}
\end{equation}
On the other hand, the rotation velocity in the equatorial plane can be calculated by dark matter density profile as well. Using the empirical dark matter density profile obtained from numerical simulations, we can then obtain the space-time metric assuming $f(r)=g(r)$ (If higher order of potential effect of the dark matter is considered, then $f(r)\neq g(r)$, but the effect is very small and complex to study). This method can be applied to all dark matter density profiles, such as Cold Dark Matter (CDM) and Scalr Field Dark Matter (SFDM).

Case I: CDM. The density profile of CDM is the NFW profile obtained from numerical simulations based on CDM and $\Lambda$CDM \citep{1991ApJ...378..496D,1997ApJ...490..493N,1996ApJ...462..563N}. We have below expressions for the density profile and rotation velocity
\begin{equation}
\rho_{NFW}(r)=\dfrac{\rho_{c}}{\dfrac{r}{R_{s}}(1+\dfrac{r}{R_{s}})^{2}},~~~~V_{NFW}(r)=\sqrt{4\pi G\rho_{c}R^{3}_{s}}\sqrt{\dfrac{1}{r}[ln(1+\dfrac{r}{R_{s}})-\dfrac{r/R_{s}}{1+\dfrac{r}{R_{s}}}]},
\label{CDM3}
\end{equation}
where $\rho_{c}$ is the density of the universe at the moment when the halo collapsed and $R_{s}$ is the characteristic radius. Using the relation between the rotation velocity and the metric coefficient function (Eq. 2), we obtain $f(r)$ and $g(r)$ as
\begin{equation}
f(r)=g(r)=exp[2\int\dfrac{V^{2}_{NFW}(r)}{r}dr]=[1+\dfrac{r}{R_{s}}]^{-\dfrac{8\pi G\rho_{c}R^{3}_{s}}{c^{2}r}}.
\label{CDM4}
\end{equation}

Case II: SFDM. This model has two density profiles which are BEC profile and finite BEC profile \citep{2002CQGra..19.6259U,2011JCAP...05..022H}. In order to simplify the discussion, we focus on BEC profile. This profile corresponds to the static solution of Kiein-Gordon equation and a quadratic potential for the scalar field $\varphi$. The profile and its corresponding rotation velocity are
\begin{equation}
\rho_{SFDM}(r)=\rho_{c}\dfrac{sin(kr)}{kr},~~~~V_{SFDM}(r)=\sqrt{\dfrac{4G\rho_{c}R^{2}}{\pi}[\dfrac{sin(\pi r/R)}{\pi r/R}-cos(\dfrac{\pi r}{R})]},
\label{SFDM1}
\end{equation}
where $k$ is determined by Compton relationship, $R=\pi/k$ is the radius at which the pressure and density are zero, and $\rho_{c}$ is the central density. Using Eq. 2, we obtain
\begin{equation}
f(r)=g(r)=exp[-\dfrac{8G\rho_{c}R^{2}}{\pi}\dfrac{sin(\pi r/R)}{\pi r/R}].
\label{SFDM3}
\end{equation}

\section{SPHERICAL SYMMETRIC BLACK HOLE METRIC IN DARK MATTER HALO}
\subsection{General method}
We now consider black holes surrounded by dark matter halo. From the pure dark matter space-time (Eq. 1), we can obtain the corresponding energy-momentum tensors. Given that energy-momentum tensors in turn lead to many kinds of space-time metric, we try to find the space-time metric of black hole surrounded by dark matter halo which can reduce to Schwarzschild metric when dark matter is not included. 

In General Relativity (GR), the Einstein field equation is given by
\begin{equation}
R^{\nu}_{~\mu}-\dfrac{1}{2}\delta^{\nu}_{~\mu}R=\kappa^{2}T^{\nu}_{~\mu}.
\label{SPBH7}
\end{equation}
If $T^{\nu}_{~\mu}=diag[-\rho,p_{r},p,p]$, we can calculate the non-zero energy-momentum tensors for pure dark matter space-time metric, which are given by
\begin{equation}
\kappa^{2}T^{t}_{t}(DM)=g(r)(\dfrac{1}{r}\dfrac{g^{'}(r)}{g(r)}+\dfrac{1}{r^{2}})-\dfrac{1}{r^{2}},$$$$
\kappa^{2}T^{r}_{r}(DM)=g(r)(\dfrac{1}{r^{2}}+\dfrac{1}{r}\dfrac{f^{'}(r)}{f(r)})-\dfrac{1}{r^{2}},$$$$
\kappa^{2}T^{\theta}_{\theta}(DM)=\kappa^{2}T^{\phi}_{\phi}(DM)=\dfrac{1}{2}g(r)(\dfrac{f^{''}(r)f(r)-f^{'2}(r)}{f^{2}(r)}+\dfrac{1}{2}\dfrac{f^{'2}(r)}{f^{2}(r)}+\dfrac{1}{r}(\dfrac{f^{'}(r)}{f(r)}+\dfrac{g^{'}(r)}{g(r)})+\dfrac{f^{'}(r)g^{'}(r)}{2f(r)g(r)}).
\label{SPBH8}
\end{equation}

In order to include black hole, we treat dark matter as part of the energy-momentum tensor $T^{\nu}_{~\mu}$. Since the Schwarzschild black hole corresponds to an energy-momentum tensor of 0, then if considering Schwarzschild black hole in dark matter halo, we only need to consider the energy-momentum tensor of dark matter in the energy-momentum tensors. The space-time metric including black hole is thus given by
\begin{equation}
ds^{2}=-(f(r)+F_{1}(r))dt^{2}+(g(r)+F_{2}(r))^{-1}dr^{2}+r^{2}(d\theta^{2}+sin^{2}\theta d\phi^{2}).
\label{SPBH9}
\end{equation}
We redefine the metric coefficient functions as
\begin{equation}
F(r)=f(r)+F_{1}(r),$$$$
G(r)=g(r)+F_{2}(r).
\label{SPC10}
\end{equation}
For such dark matter-black hole system in which dark matter is considered as part of Einstein $T^{\mu}_{~~\mu}$, the Einstein field equation (Eq. 7) becomes 
\begin{equation}
R^{\nu}_{~\mu}-\dfrac{1}{2}\delta^{\nu}_{~\mu}R=\kappa^{2}T^{\nu}_{~\mu}=\kappa^{2}(T^{\nu}_{~\mu}+T^{\nu}_{~\mu}(DM)).
\label{SPBH11}
\end{equation}
Inserting the space-time metric (Eq. 9) into the new Einstein field equation (Eq. 11), we obtain
\begin{equation}
(g(r)+F_{2}(r))(\dfrac{1}{r^{2}}+\dfrac{1}{r}\dfrac{g^{'}(r)+F^{'}_{2}(r)}{g(r)+F_{2}(r)})=g(r)(\dfrac{1}{r^{2}}+\dfrac{1}{r}\dfrac{g^{'(r)}}{g(r)}),$$$$
(g(r)+F_{2}(r))(\dfrac{1}{r^{2}}+\dfrac{1}{r}\dfrac{f^{'}(r)+F^{'}_{1}(r)}{f(r)+F_{1}(r)})=g(r)(\dfrac{1}{r^{2}}+\dfrac{1}{r}\dfrac{f^{'(r)}}{f(r)}).
\label{SPBH12}
\end{equation}
where $F_{1}(r)$ and $F_{2}(r)$ are given by
\begin{equation}
(rg(r)+rF_{2}(r))F^{'}_{2}(r)+(rg^{'}(r)+g(r)-r\dfrac{g^{'}(r)}{g(r)})F_{2}(r)+F^{2}_{2}(r)=rg^{'}(r)-rg(r)g^{'}(r),$$$$
\dfrac{f^{'}(r)+F^{'}_{1}(r)}{f(r)+F_{1}(r)}=\dfrac{g(r)}{g(r)+F_{2}(r)}(\dfrac{1}{r}+\dfrac{f^{'}(r)}{f(r)})-\dfrac{1}{r}.
\label{SPBH13}
\end{equation}
The first equation of Eq. (13) is a second Abel equation which is resolvable in our case to obtain the expressions of $F_{1}(r)$ and $F_{2}(r)$ as
\begin{equation}
F_{1}(r)=exp[\int \dfrac{g(r)}{g(r)+F_{2}(r)}(\dfrac{1}{r}+\dfrac{f^{'}(r)}{f(r)})-\dfrac{1}{r} dr]-f(r),$$$$
F_{2}(r)=-\dfrac{2GM}{c^{2}r}.
\label{SPBH14}
\end{equation}

Then the black hole space-time metric in dark matter halo is given by
\begin{equation}
ds^{2}=-exp[\int \dfrac{g(r)}{g(r)-\dfrac{2GM}{c^{2}r}}(\dfrac{1}{r}+\dfrac{f^{'}(r)}{f(r)})-\dfrac{1}{r} dr]dt^{2}+(g(r)-\dfrac{2GM}{c^{2}r})^{-1}dr^{2}
+r^{2}(d\theta^{2}+sin^{2}\theta d\phi^{2}).
\label{SPBH15}
\end{equation}

If we do not consider dark matter halo, i.e., $f(r)=g(r)=1$, the indefinite integral will become constant and we have
\begin{equation}
F_{1}(r)+f(r)=exp[\int \dfrac{g(r)}{g(r)+F_{2}(r)}(\dfrac{1}{r}+\dfrac{f^{'}(r)}{f(r)})-\dfrac{1}{r} dr]=1-\dfrac{2GM}{c^{2}r},$$$$
F_{2}(r)+g(r)=\dfrac{1}{r}\int[-\dfrac{g^{'}(r)}{g^{2}(r)}+\dfrac{g^{'}(r)}{g(r)}-g^{'}(r)+\dfrac{g(r)}{r}]rdr-\dfrac{2GM}{c^{2}r}=1-\dfrac{2GM}{c^{2}r}.
\label{SPBH16}
\end{equation}
In this case, the space-time reduces to Schwarzschild black hole space-time. Therefore the space-time Eq.(15) describes Schwarzschild black hole surrounded by dark matter halo. For any given dark matter density profile, we can obtain the corresponding space-time in this way.

\subsection{Cold dark matter (CDM)}
For CDM dark matter halo, if $f(r)=g(r)$, we obtain $F_{1}(r)=F_{2}(r)=-\dfrac{2GM}{rc^{2}}$, the black hole space-time metric coefficient functions are given by
\begin{equation}
F_{1}(r)+f(r)=F_{2}(r)+g(r)=[1+\dfrac{r}{R_{s}}]^{-\dfrac{8\pi G\rho_{c}R^{3}_{s}}{c^{2}r}}- \dfrac{2GM}{rc^{2}}.
\label{CDM17}
\end{equation}
The corresponding black hole space-time is
\begin{equation}
ds^{2}=-[[1+\dfrac{r}{R_{s}}]^{-\dfrac{8\pi G\rho_{c}R^{3}_{s}}{c^{2}r}}- \dfrac{2GM}{rc^{2}}]dt^{2}+([1+\dfrac{r}{R_{s}}]^{-\dfrac{8\pi G\rho_{c}R^{3}_{s}}{c^{2}r}}- \dfrac{2GM}{rc^{2}})^{-1}dr^{2}
+r^{2}(d\theta^{2}+sin^{2}\theta d\phi^{2}).
\label{CDM11}
\end{equation}

\subsection{Scalar field dark matter (SFDM)}
For SFDM halo, if $f(r)=g(r)$, we obtain $F_{1}(r)=F_{2}(r)=-\dfrac{2GM}{rc^{2}}$, the black hole space-time metric coefficient functions are given by
\begin{equation}
F_{1}(r)+f(r)=F_{2}(r)+g(r)=exp[-\dfrac{8G\rho_{c}R^{2}}{\pi}\dfrac{sin(\pi r/R)}{\pi r/R}]-\dfrac{2GM}{rc^{2}}.
\label{SFDM10}
\end{equation}
The corresponding black hole space-time is
\begin{equation}
ds^{2}=-[exp[-\dfrac{8G\rho_{c}R^{2}}{\pi}\dfrac{sin(\pi r/R)}{\pi r/R}]-\dfrac{2GM}{rc^{2}}]dt^{2}+
(exp[-\dfrac{8G\rho_{c}R^{2}}{\pi}\dfrac{sin(\pi r/R)}{\pi r/R}]-\dfrac{2GM}{rc^{2}})^{-1}dr^{2}$$$$
+r^{2}(d\theta^{2}+sin^{2}\theta d\phi^{2}).
\label{SFDM11}
\end{equation}

\section{ROTATIONAL BLACK HOLE METRIC IN DARK MATTER HALO}
We now generalize spherical symmetric black hole to rotational black hole surrounded by dark matter halo basing on the Newman-Jains method \citep[e.g.,][]{1965JMP.....6..915N,2014PhRvD..90f4041A,2015arXiv151201498T}. Our work directly follow the procedure in \cite{2015arXiv151201498T}.

In Newman-Jains method, $\Sigma^{2}=r^{2}+a^{2}cos^{2}\theta$. The rotational black hole space-time metric surrounded by dark matter halo is
\begin{equation}
ds^{2}=-(1-\dfrac{r^{2}-G(r)r^{2}}{\Sigma^{2}})dt^{2}+\dfrac{\Sigma^{2}}{\Delta}dr^{2}+\dfrac{2(r^{2}-G(r)r^{2})a sin^{2}\theta}{\Sigma^{2}}d\phi dt+$$$$
\Sigma^{2}d\theta^{2}+\dfrac{sin^{2}\theta}{\Sigma^{2}}((r^{2}+a^{2})^{2}-a^{2}\Delta sin^{2}\theta)d\phi^{2},
\label{KBH1}
\end{equation}
where
\begin{equation}
\Delta=r^{2}G(r)+a^{2}=r^{2}G(r)+(\dfrac{J}{Mc})^{2},$$$$
G(r)=F(r)=f(r)+F_{1}(r)=g(r)+F_{2}(r).
\label{KBH2}
\end{equation}
Here below we present the explicit expressions of the space-time metric for the dark matter models considered in this work.

Case I: CDM
\begin{equation}
ds^{2}=-(1-\dfrac{r^{2}+\dfrac{2GMr}{c^{2}}-r^{2}[1+\dfrac{r}{R_{s}}]^{-\dfrac{8\pi G\rho_{c}R^{3}_{s}}{c^{2}r}}}{\Sigma^{2}})dt^{2}+\dfrac{\Sigma^{2}}{\Delta}dr^{2}+
\Sigma^{2}d\theta^{2}+\dfrac{sin^{2}\theta}{\Sigma^{2}}((r^{2}+a^{2})^{2}-a^{2}\Delta sin^{2}\theta)d\phi^{2}$$$$
+\dfrac{2(r^{2}+\dfrac{2GMr}{c^{2}}-r^{2}[1+\dfrac{r}{R_{s}}]^{-\dfrac{8\pi G\rho_{c}R^{3}_{s}}{c^{2}r}})a sin^{2}\theta}{\Sigma^{2}}d\phi dt,$$$$
\Delta=r^{2}[1+\dfrac{r}{R_{s}}]^{-\dfrac{8\pi G\rho_{c}R^{3}_{s}}{c^{2}r}}-\dfrac{2GMr}{c^{2}}+a^{2}.
\label{CDM30}
\end{equation}

Case II: SFDM
\begin{equation}
ds^{2}=-(1-\dfrac{r^{2}+\dfrac{2GMr}{c^{2}}-r^{2}exp[-\dfrac{8G\rho_{c}R^{2}}{\pi}\dfrac{sin(\pi r/R)}{\pi r/R}]}{\Sigma^{2}})dt^{2}+\dfrac{\Sigma^{2}}{\Delta}dr^{2}
+\Sigma^{2}d\theta^{2}$$$$
+\dfrac{2(r^{2}+\dfrac{2GMr}{c^{2}}
-r^{2}exp[-\dfrac{8G\rho_{c}R^{2}}{\pi}\dfrac{sin(\pi r/R)}{\pi r/R}])a sin^{2}\theta}{\Sigma^{2}}d\phi dt
+\dfrac{sin^{2}\theta}{\Sigma^{2}}((r^{2}+a^{2})^{2}-a^{2}\Delta sin^{2}\theta)d\phi^{2},~~~$$$$
\Delta=r^{2}exp[-\dfrac{8G\rho_{c}R^{2}}{\pi}\dfrac{sin(\pi r/R)}{\pi r/R}]-\dfrac{2GMr}{c^{2}}+a^{2}.
\label{SFDM30}
\end{equation}

Eq. (23) and Eq. (24) represent the space-time metric of rotational black holes surrounded by CDM halo and SFDM halo, respectively. If we do not consider the dark matter halo, they will reduce to the Kerr black hole space-time. In addition, we have calculated the energy-momentum tensors of Eq. (23) and Eq. (24) and verified that they can reduce to the case of spherical symmetry (Eq. 18 and Eq. 20). Therefore our black hole spacetime metric of rotating black hole (Eq. 23 and Eq. 24) satisfy the Einstein field equation. These results would be useful to study dark matter near Kerr black holes, especially for high spin black holes.

%\section{PROPERTIES OF BLACK HOLESS SURROUND BY DARK MATTER HALO}
\section{DISCUSSION}
We now discuss the behaviour of dark matter profiles and properties of black holes for the rotational black hole space-time metric surrounded by dark matter halo that we obtained in Section 4. Our dark matter halo parameters come from the Low Surface Brightness (LSB) galaxy ESO1200211 presented in \cite{2017arXiv170806681F} and \cite{2012MNRAS.422..282R}. We adopted, for CDM, $\rho_{c}=2.45*10^{-3}M_{\circledcirc}/pc^{3}$, $R_{s}=5.7Kpc$ and for SFDM, $\rho_{c}=13.66*10^{-3}M_{\circledcirc}/pc^{3}$, $R=2.92Kpc$.

\subsection{The cusp phenomenon}
The NFW density profile is obtained in numerical simulations of CDM and $\Lambda$CDM. When the distance $r$ from the black hole is below 1-2 kpc, this profile exhibits the ``cusp" phenomenon \citep{2010AdAst2010E...5D}. But for the SFDM density profile, no ``cusp" phenomenon appears and the density is close to be constant at small distance \citep{2012MNRAS.422..282R}. In this work, we investigate the dark matter density profile by taking into account the relativistic effects (black hole).

From the Einstein field equation (Eq. 11), the energy density $\rho$ in dark matter halo is given by
\begin{equation}
\kappa^{2}\rho=\dfrac{1}{r^{2}}-G(r)(\dfrac{1}{r}\dfrac{G^{'}(r)}{G(r)}+\dfrac{1}{r^{2}}).
\label{SPIKE1}
\end{equation}
Because the motion velocity of dark matter particle is much smaller than the speed of light, the energy density of dark matter then approximates to the mass density. On the other hand, the density profile is usually observed at large distances from the black hole, so that the black hole spin $a$ then can be approximated as zero. Here below we show the explicit expressions for the dark matter models considered in this work.

Case I: CDM
\begin{equation}
\kappa^{2}\rho=\dfrac{1}{r^{2}}(1-[1+\dfrac{r}{R_{s}}]^{-\dfrac{8\pi G\rho_{c}R^{3}_{s}}{c^{2}r}})-\dfrac{1}{r}[1+\dfrac{r}{R_{s}}]^{-\dfrac{8\pi G\rho_{c}R^{3}_{s}}{c^{2}r}}[\dfrac{8\pi G\rho_{c}R^{3}_{s}}{c^{2}r^{2}}ln(1+\dfrac{r}{R_{s}})-\dfrac{8\pi G\rho_{c}R^{3}_{s}}{c^{2}r(r+R_{s})}].
\label{CDM20}
\end{equation}

Case II: SFDM
\begin{equation}
\kappa^{2}\rho=\dfrac{1}{r^{2}}(1-exp[-\dfrac{8G\rho_{c}R^{2}}{\pi}\dfrac{sin(\pi r/R)}{\pi r/R}])-
\dfrac{1}{r} \dfrac{8G\rho_{c}R^{2}}{\pi}exp[-\dfrac{8G\rho_{c}R^{2}}{\pi}\dfrac{sin(\pi r/R)}{\pi r/R}]
[\dfrac{1}{r}cos(\dfrac{\pi r}{R})-\dfrac{R}{\pi r^{2}}sin(\dfrac{\pi r}{R})].
\label{SFDM21}
\end{equation}

\begin{figure}[htbp]
  \centering
  \includegraphics[scale=0.4]{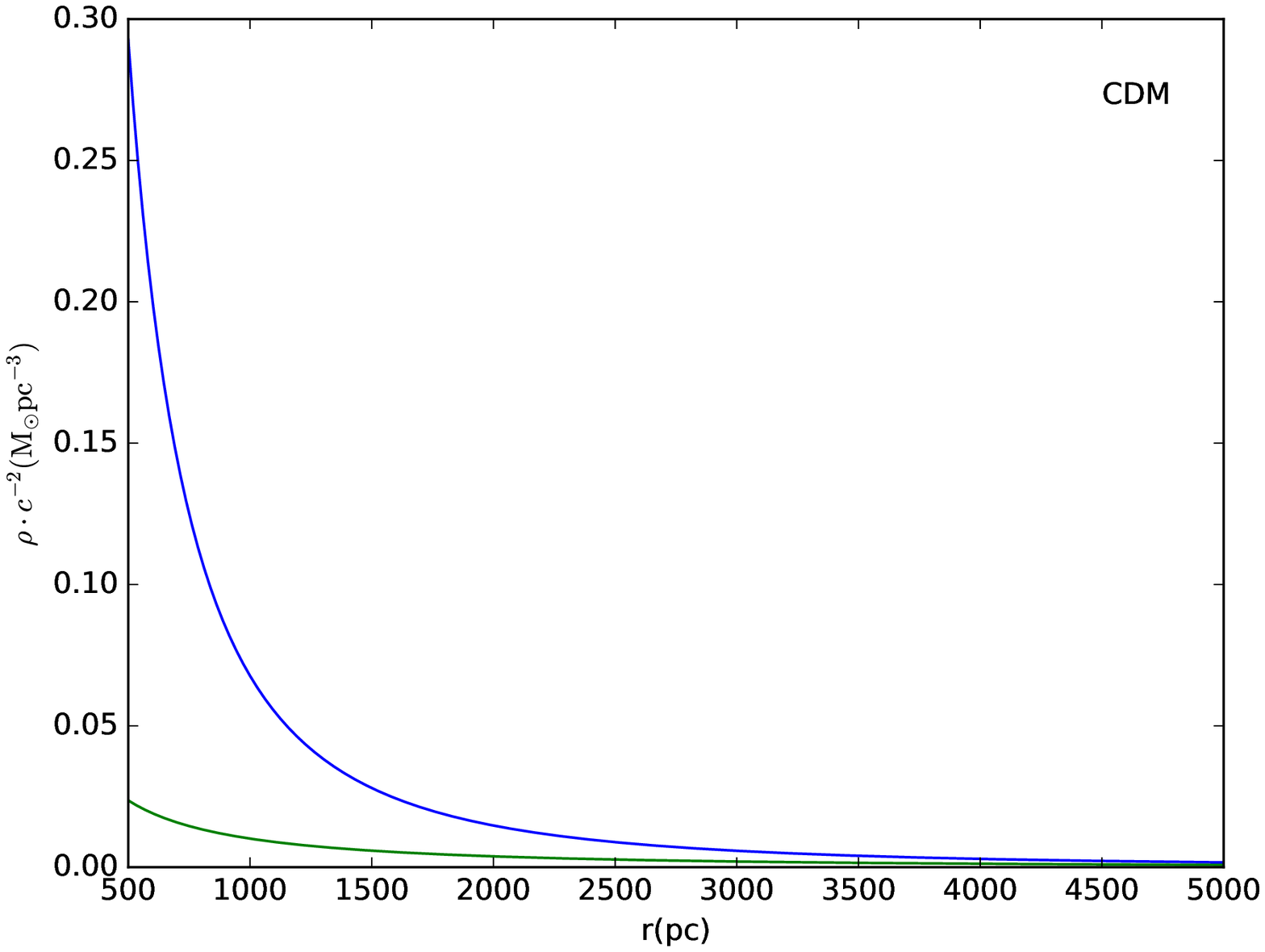}
  \includegraphics[scale=0.4]{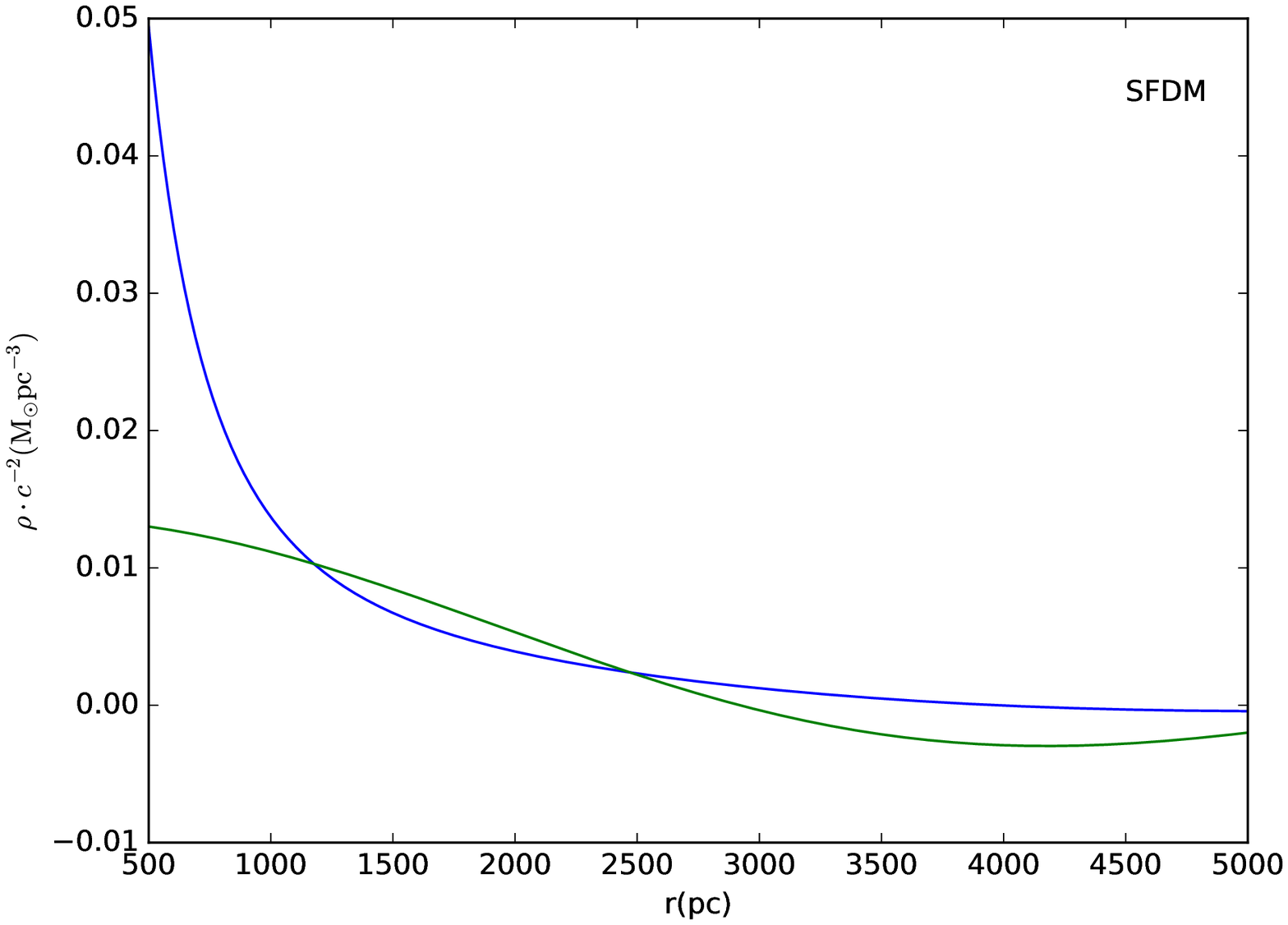}
   \caption{CDM profiles (left figure) and SFDM profiles (right figure). The green and blue lines represent the dark matter density profile without and with general relativity considered, respectively.}
  \label{fig:1}
\end{figure}

From Eq. 26 and Eq. 27, it is clear that the energy density $\rho$ reaches to infinity when the distance $r$ is close to 0. Fig. 1 shows the profiles of CDM and SFDM models, for both the cases of with and without relativist effects considered. We find that when relativist effects are taking into account, both models show ``cusp" and the energy density $\rho$ is significantly enhanced close to the black hole, especially for the SFDM profile. For the CDM model, the ``cusp" still exists even if we do not consider relativistic effects, while for the SFDM model, there is no ``cusp" when relativistic effect is not considered. This is an interesting discover for dark matter density profiles.

\subsection{Black hole properties in dark matter halo}
Basing on the space-time metric (Eq. 23, 24), we discuss now qualitatively how the dark matter halo changes the Kerr black hole properties which include horizon, ergosphere and singularity.

Horizon: Like Kerr space-time, the rotational black hole surrounded by dark matter halo has two horizons, i.e., the inner horizon and the event horizon. Because dark matter does not make black hole produce new horizon, the horizon is defined by $\Delta=0$. Through numerical calculations, we obtain that the dark matter halo makes the horizon to increase by about $10^{-7}$ orders of magnitude. Comparing the two kinds of dark matter models considered in this work, we find that the change of horizon is larger for CDM and smaller for SFDM. The size of two horizons depends on the parameters of dark matter halo, i.e., the scale density $\rho_{c}$ and characteristic radius $R_{c}$ or the scalar curvature.

Ergosphere: The ergosphere exists between the event horizon and the inner static limit surface, which is defined by $g_{tt}=0$. Through numerical calculations, we obtain the following results. Firstly, the dark matter makes the size of ergosphere to decrease by about $10^{-7}$ orders of magnitude, and the energy extraction (Penrose Process) of black hole will decrease. Secondly, different dark matter models have different effects on the ergosphere. The change of ergosphere is larger for CDM and smaller for SFDM.

Singularity: Through calculating the scalar curvature, we find that these black hole space-time are singular at $\Sigma^{2}=r^{2}+a^{2}cos^{2}\theta=0$ which represents a ring in Boyer-Lindquist coordinates. This implies that dark matter can not change the singularity of black hole. Yet, our results are based on $f(r)=g(r)$.

\section{SUMMARY}
%For the first time, we obtain the analytical form of the space-time metric of black hole in dark matter halo in stationary situation. Using the relation between the rotation velocity (in the equatorial plane) and spherically symmetric space-time metric coefficient $f(r)$ (suppose $f(r)=g(r))$, we first obtain the space-time metric for pure dark matter density profile. By considering the dark matter halo in spherically symmetric space-time as part of the energy-momentum tensor $T^{\mu}_{~\nu}$ in Einstein field equation, we then obtain the space-time in spherically symmetric black holes in dark matter halo. Utilizing Newman-Jains method, we finally generalize spherically symmetric black holes to rotational black holes for the CDM and SFDM dark matter models and discuss qualitatively the behaviour of dark matter density profile and black hole properties. Firstly, we find that all dark matter halo models near the black hole center will produce the "cusp" phenomenon, especially for SFDM. Secondly, through numerical calculation, we analyze how dark matter halo changes the black hole space-time, including black hole horizon, ergosphere and singularity. In general, dark matter halo makes the value of horizon to increase and the size of ergosphere to decrease, but the change is minor. Dark matter does not change the singularity. For galaxies with different parameters of $\rho$ and $R_{c}$, the above results are different.

In this work, we obtain the analytical form of black hole space-time in dark matter halo in stationary situation, for the first time. We start from the pure dark matter space-time metric and then derive the one for spherical symmetric black holes surrounded by dark matter halo. Finally we generalize spherical symmetric black hole to rotational black hole in dark matter halo. Qualitative analysis, using LSB galaxy data, of the behaviour of dark matter density profile and properties of black hole for the CDM and SFDM halos considered in this work show that both dark matter density profiles near the black hole produce the  ``cusp" phenomenon, and both makes the black hole horizon to increase and the ergosphere to decrease, though the magnitude is small. On the other hand, dark matter does not change the singularity of black holes.

Our results are useful to study the interaction of black hole and dark matter halo in stationary situation. Particularly, the black hole can enhance the dark matter profile significantly at small but observable distances producing the ``cusp" phenomenon. This phenomenon could be tested from observations of the Milky Way. 

In the future work, we will try to investigate applications of our findings in astrophysics, such as to study the TDE of black holes surrounded by dark matter, black hole shadow in dark matter halo and dynamical processes of black hole affected by dark matter, etc.


\begin{thebibliography}{99}

\bibitem[Balberg \& Shapiro(2002)]{2002PhRvL..88j1301B} Balberg, S., \& Shapiro, S.~L.\ 2002, Physical Review Letters, 88, 101301

\bibitem[Balberg et al.(2002)]{2002ApJ...568..475B} Balberg, S., Shapiro, S.~L., \& Inagaki, S.\ 2002, \apj, 568, 475

\bibitem[Fay(2004)]{2004A&A...413..799F} Fay, S.\ 2004, \aap, 413, 799

\bibitem[Matos et al.(2005)]{2005GReGr..37..769M} Matos, T., N{\'u}{\~n}ez, D., \& Sussman, R.~A.\ 2005, General Relativity and Gravitation, 37, 769

\bibitem[Matos \& Nunez(2003)]{2003astro.ph..3594M} Matos, T., \& Nunez, D.\ 2003, arXiv:astro-ph/0303594

\bibitem[Gondolo \& Silk(1999)]{1999PhRvL..83.1719G} Gondolo, P., \& Silk, J.\ 1999, Physical Review Letters, 83, 1719

\bibitem[Sadeghian et al.(2013)]{2013PhRvD..88f3522S} Sadeghian, L., Ferrer, F., \& Will, C.~M.\ 2013, \prd, 88, 063522

\bibitem[Fields et al.(2014)]{2014PhRvL.113o1302F} Fields, B.~D., Shapiro, S.~L., \& Shelton, J.\ 2014, Physical Review Letters, 113, 151302

\bibitem[de Blok(2010)]{2010AdAst2010E...5D} de Blok, W.~J.~G.\ 2010, Advances in Astronomy, 2010, 789293

\bibitem[Dubinski \& Carlberg(1991)]{1991ApJ...378..496D} Dubinski, J., \& Carlberg, R.~G.\ 1991, \apj, 378, 496

\bibitem[Navarro et al.(1997)]{1997ApJ...490..493N} Navarro, J.~F., Frenk, C.~S., \& White, S.~D.~M.\ 1997, \apj, 490, 493

\bibitem[Navarro et al.(1996)]{1996ApJ...462..563N} Navarro, J.~F., Frenk, C.~S., \& White, S.~D.~M.\ 1996, \apj, 462, 563

\bibitem[Ure{\~n}a-L{\'o}pez et al.(2002)]{2002CQGra..19.6259U} Ure{\~n}a-L{\'o}pez, L.~A., Matos, T., \& Becerril, R.\ 2002, Classical and Quantum Gravity, 19, 6259

\bibitem[Harko(2011)]{2011JCAP...05..022H} Harko, T.\ 2011, Journal of Cosmology and Astroparticle Physics, 5, 022

\bibitem[Begeman et al.(1991)]{1991MNRAS.249..523B} Begeman, K.~G., Broeils, A.~H., \& Sanders, R.~H.\ 1991, \mnras, 249, 523

\bibitem[Azreg-A{\"i}nou(2014)]{2014PhRvD..90f4041A} Azreg-A{\"i}nou, M.\ 2014, \prd, 90, 064041

\bibitem[Newman \& Janis(1965)]{1965JMP.....6..915N} Newman, E.~T., \& Janis, A.~I.\ 1965, Journal of Mathematical Physics, 6, 915

\bibitem[Toshmatov et al.(2015)]{2015arXiv151201498T} Toshmatov, B., Stuchl{\'{\i}}k, Z., \& Ahmedov, B.\ 2015, arXiv:1512.01498

\bibitem[Fernandez-Hernandez et al.(2017)]{2017arXiv170806681F} Fernandez-Hernandez, L.~M., Rodriguez-Meza, M.~A., \& Matos, T.\ 2017, arXiv:1708.06681 

\bibitem[Robles \& Matos(2012)]{2012MNRAS.422..282R} Robles, V.~H., \& Matos, T.\ 2012, \mnras, 422, 282 






\end{thebibliography}
\end{document}